# Resonant tunnelling in a quantum oxide superlattice

Woo Seok Choi[1,2], Sang A Lee[2,3], Jeong Ho You[4], Suyoun Lee[1,5], and Ho Nyung Lee[1*]

[1]Materials Science and Technology Division, Oak Ridge National Laboratory, Oak Ridge, TN 37831, USA

[2]Department of Physics, Sungkyunkwan University, Suwon, Gyeonggi-do 440-746, Korea

[3]Institute of Basic Science, Sungkyunkwan University, Suwon, Gyeonggi-do 440-746, Korea

[4]Department of Mechanical Engineering, Southern Methodist University, Dallas, TX 75205, USA

[5]Electronic Materials Research Center, Korea Institute of Science and Technology, Seoul 136-791, Korea

[*]e-mail: hnlee@ornl.gov.

Resonant tunnelling is a quantum mechanical process that has long been attracting both scientific and technological attention owing to its intriguing underlying physics and unique applications for high-speed electronics. The materials system exhibiting resonant tunnelling, however, has been largely limited to the conventional semiconductors, partially due to their excellent crystalline quality. Here we show that a deliberately designed transition metal oxide superlattice exhibits a resonant tunnelling behaviour with a clear negative differential resistance. The tunnelling occurred through an atomically thin, lanthanum $\delta$-doped $SrTiO_3$ layer, and the negative differential resistance was realized on top of the bi-polar resistance switching typically observed for perovskite oxide junctions. This combined process resulted in an extremely large resistance ratio (~$10^5$) between the high and low resistance states. The unprecedentedly large control found in atomically thin $\delta$-doped oxide superlattices can open a door to novel oxide-based high-frequency logic devices.

Introduction

In quantum well (QW) heterostructures, the probability of quantum mechanical tunnelling depends on the available quantized states at both the originating and the receiving sides of the junction. Therefore, the tunnelling current is usually not a monotonically increasing function with respect to the external bias. In particular, quantized resonant states can form in a QW between the two barriers, and can give rise to an interesting tunnelling behaviour, so called, resonant tunnelling (RT). Here, the energy of the wavefunctions with discrete levels can shift and align by applying an external bias (Fig. 1)[1,2]. When properly aligned, the tunnelling current peaks, exhibiting a negative differential resistance (NDR) just above the resonant bias ($V_R$). Such an intriguing NDR behaviour can be exploited in various devices such as tunnel diodes and RT transistors. More importantly, the RT phenomena offer unique insight into electrical transport properties of materials, such as localized defect states, collective electronic excitations, and QW band structures. Up to date, NDR and accompanying RT behaviours have been studied mostly in Si-based or III-V compound semiconductors[1-5]. Excellent crystalline quality of the conventional semiconductors and actual realization of two dimensional heterostructures with nanometre-thick barrier layers enabled the observation of the RT behaviour.

Based on recent advances in oxide thin film synthesis, precision design of complex oxide thin films and heterostructures by pulsed laser epitaxy (PLE) or molecular beam epitaxy (MBE) at the atomic scale has become available[6-8]. Indeed, many previous studies demonstrated emergent physical phenomena in conjunction with the quantum mechanical tunnelling in transition metal oxide (TMO)-based tunnel junctions, utilizing the exotic properties, including superconductivity[9], magnetoresistance[10,11], electroresistance[12,13], and multiferroicity[14]. The strong coupling among charge, spin, lattice, and orbital degrees of freedom has been investigated in terms of the various tunnelling behaviours, such as Josephson effect and spin-polarized tunnelling. More recently, a large electroresistance was also observed in ferroelectric tunnel junctions[12]. The tunnelling current across the ferroelectric layer is found to be efficiently controlled by not only the polarization direction, but also, more importantly, the interfacial electronic phase of the electrode layers[13]. Unfortunately, however, a clear RT behaviour has not been experimentally realized in TMO heterostructures.

Although a clear RT behaviour has not been widely explored with TMOs, the NDR behaviour itself has been reported in various complex oxide junction structures[15-18], In particular, nonvolatile resistance switching (RS) in TMO-based heterostructures has attracted scientific attention for developing next-generation memory devices based on memristors[19-21]. In typical RS devices, the electric resistance of a junction can be modulated by applying external bias, and the device can have two (or more) resistance states, *i.e.*, a high resistance state (HRS) and a low resistance state (LRS). The switching from a LRS to a HRS usually accompanies a NDR behaviour, as the amount of current decreases upon increasing the bias above the switching voltage.

In the following, we investigate the junction transport property of a QW superlattice (SL) precisely designed by inserting atomically thin $LaTiO_3$ (LTO) between $SrTiO_3$ (STO) barrier layers. A clear RT behaviour with NDR is observed. Moreover, as oxide heterostructures can reveal RS, we further propose that the combination of RS with NDR is an efficient way to maximize the resistance ratio between the HRS and LRS in oxide QW heterostructures.

Results

**Precision design of oxide quantum heterostructures by pulsed laser epitaxy.** In order to realize a TMO QW heterostructure, a La $\delta$-doped STO SL with atomically sharp interfaces was fabricated[22]. We believe that a sample with an excellent crystalline quality is highly necessary for the observation of the quantum resonant behaviour in TMOs. Therefore, we epitaxially grew a high quality $[LTO_1/STO_6]_{10}$ SL with minimized defect states using our atomic scale synthesis capability[6, 8]. Detailed information on the sample growth by PLE and experimental detail, as well as the excellent crystallinity of our SL is demonstrated in the Methods and Supplementary Figure 1.

A schematic drawing of the SL structure is shown in Fig. 1(a). It should be noted that La $\delta$-doping creates a two dimensional electron gas (2DEG) with high-density conducting carriers at the interface (0.5 electrons per unit cell), due to the leakage of the $3d^1$ electrons in $La^{3+}Ti^{3+}O^{2-}_3$ into the $3d^0$ states in $Sr^{2+}Ti^{4+}O^{2-}_3$[8, 23-25]. The partial $3d$ electrons, or the distribution of the $Ti^{3+}$ ions are confined around the LTO layer with the QW width of a few (5~6) unit cells [23, 26]. On the other hand, it has been shown that the STO layer remains largely insulating and can serve as a barrier layer if the thickness is larger than three unit cells (u.c.)[23, 26]. Therefore, the 2DEG formed at the interface can be considered as a QW with a thickness of a few nanometres, as shown in Fig. 1b. While many previous studies have focused on the in-plane electronic transport behaviours of such a 2DEG in titanate based oxide heterostructures[25, 27, 28], little is known about the out-of-plane junction properties[29]. In addition, unlike typical RT devices composed of tens of nanometres-thick semiconductor QWs, the tunnelling in our devices occurs through only an atomically thin layer, the fundamental lattice unit, providing a promising outlook for an ultra-fast transit time.

**Nonlinear current-voltage characteristics and negative differential resistance.** Figure 2 shows the junction current-voltage characteristics of the TMO QWs at various temperatures ($T$). At room $T$, a weak hysteresis curve, which manifests the typical bipolar RS behaviour, was observed. The two different resistance states could be achieved by switching the polarity of the bias voltage. Indeed, the memory characteristics have been confirmed to show a typical RS behaviour. In general, understanding the switching mechanisms is a major challenge in achieving better nonvolatile memory performance. In order to explain the bipolar RS phenomenon, various electric field polarity dependent models, including ion migration[19], Mott transition[30], and formation of Schottky barrier[31-33], have been suggested. In particular, the Schottky barrier formed at the interface has been identified to trigger the RS behaviour for Nb:STO based junctions[31, 33]. More specifically, changes in the Schottky barrier by either a charge trap at the defect states or oxygen migration due to applied bias voltage has been frequently attributed for the bipolar RS behaviour[31, 32, 34]. Similar mechanism, if not the same, seems to play a major role in our SL as well. As a qualitatively similar bipolar RS behaviour was also observed for a Pt/Nb:STO junction (inset of Fig. 2a), we believe that the RS behaviour originates mainly from the interface between the SL and Nb:STO.

As $T$ decreases, the overall current level increases for both voltage polarities, and the RS becomes more distinct with a larger discrepancy between the LRS and HRS. Such $T$-dependent increase of the current level can be attributed to the largely enhanced carrier mobility and dielectric constant of STO at low $T$[25]. In particular, similar $T$-dependence of the current level for the Pt/Nb:STO junction (inset of Fig. 2a) suggests that the bottom electrode Nb:STO is largely responsible for the increasing current level with decreasing $T$. At $T < \sim$220 K, an interesting transport behaviour starts to develop at ~1.2 V while maintaining the overall bipolar RS feature of our SL. A clear and smooth NDR behaviour is then observed by further ramping up the voltage to the positive bias direction. It should be noted that such a smooth peak in $I(V)$ curve is significantly different from the usual abrupt NDR features observed for conventional RS in TMOs. For general RS (for both bipolar and unipolar), the current level drops immediately when the switching occurs from a LRS to a HRS, in stark contrast to what we observe here. Moreover, a further increase in the current above the clear peak feature in $I(V)$ curve was observed, which has never been reported for conventional RS, to the best of our knowledge. These characteristic features signify that the origin of NDR in our QW SLs is fundamentally different from that of the conventional RS often observed in oxide thin films.

It is also important to point out that the NDR behaviour uniquely originates from the TMO QW structure. $I(V)$ curves for a Pt/Nb:STO junction as a function of $T$ shown in the inset of Fig. 2a represent only a typical bipolar RS behaviour of Nb:STO junctions. While similar RS behaviours have been observed for various Nb:STO based junctions as discussed previously[32, 33], the smooth peak feature and clear NDR behaviour were absent, indicating that the deliberately designed TMO SL with QWs is indeed responsible for the unique feature. Moreover, upon differentiating the *I*(*V*) curves of the SLs as shown in Fig. 2b, we indeed find two clear zero-crossing points for $T < 200$ K. This result again indicates that the NDR is a consequence of the resonant states formed inside the QWs, rather than associated with the conventional RS. Observation of NDR only at lower *T* possibly due to the reduced phonon scattering further confirms the RT behaviour.

**Theoretical calculations on resonant tunnelling.** To verify that the NDR behaviour indeed originates from RT, we have performed a theoretical analysis by taking into account the band gap, effective mass, and dielectric constant ($\epsilon_r$) of the TMO heterostructure (see Method section for detail). Among these parameters, $\epsilon_r$ of an oxide heterostructure with 2DEGs has not been exactly understood. However, it is expected that the creation of interfacial charges would drastically reduce the $\epsilon_r$ value of the highly dielectric STO. As this change leads to a significant modification in the dielectric screening of the conducting carriers, we focus here on the effect of $\epsilon_r$. Figure 3 shows the evolution of energy levels of the quantized states and $V_R$ for [LTO$_1$/STO$_6$]$_{10}$ SL as a function of $\epsilon_r$. The quantized energy levels show the values in the absence of external electric field. The actual potential wells and wavefunctions within the well with corresponding energy levels are represented in Fig. 1b for $\epsilon_r = 100$. As $\epsilon_r$ increases, the well becomes shallower, weakening the quantum confinement — *i.e.*, both the energy difference between the states and the number of confined states are reduced upon the increase of $\epsilon_r$. For $\epsilon_r \leq 100$, the well is deep enough to accommodate three confined states (ground, first and second excited states) within the QW, while only two confined states are possible for $\epsilon_r > 100$ (Fig. 3a). Based on the energy level separation and the QW geometry, we can calculate $V_R$ as a function of $\epsilon_r$ (Fig. 3b). As $\epsilon_r$ increases, a smaller $V_R$ is expected because of the smaller energy difference between the ground and first excited states. For $\epsilon_r = 100$, $V_R = 1.23$ V is needed to induce RT between the ground and first excited states in the deliberately designed oxide SL, as shown in Fig. 1c. The theoretical calculation implies that it would be rather difficult to observe a well-defined higher order peak in the tunnelling current. The confinement of the second excited state is only possible for $\epsilon_r <$ 100 ideally. Even then, the state becomes easily unbound by applying a small electric field. Practically, the overall high current level makes it impossible to observe clear high voltage features. Especially, the current substantially increased to a very high level at low $T$ and the current compliance set to prevent thermal damage of the oxide QW obscures the peak feature. (We noted some $T$ fluctuation near the lowest $T$ due to the Joule heating, but the $I(V)$ characteristic was highly stable and reproducible.)

Discussion

It should be noted that the RT behaviour is observed for the positive bias only, which can be attributed to the asymmetric sample geometry and resultant current profile. The current level for the negative bias is too large (for both HRS and LRS), possibly hindering the observation of the NDR behaviour. In addition, the current reaches to the compliance limit much faster, which makes it even more difficult to observe the NDR behaviour for the negative bias. Finally, the $I(V)$ characteristic of our oxide device is remarkably similar to that from a $SiO_2$/Si resonant tunnelling diode,[4] strongly supporting that the NDR behaviour is based on RT in our TMO heterostructure.

We note that a few previous studies on TMO junctions claimed an observation of resonant states[16, 35]. These reports indicated that resonant states could be formed with $V_R$ on the order of 0.1 ~ 1.0 V due to defects or impurity states. The observations are based on the chemical doping, *i.e.*, oxygen vacancies in $SrTiO_{3-\delta}$ and Mn doping into $SrRuO_3$ ($SrRu_{0.95}Mn_{0.05}O_3$) for Pt/$SrTiO_{3-\delta}$ and $SrRu_{0.95}Mn_{0.05}O_3$/Nb:STO interfaces, respectively. On the other hand, our discovery of clear RT is based on a deliberately designed TMO QW SL, which might provide an unprecedented opportunity to develop novel tunnelling oxide electronic devices.

The observation of RT behaviour further provides quantitative information useful for understanding the physics of RT phenomena and insight for technological applications. We note that such a quantitative analysis could not have been conducted on TMO QWs, as RT behaviour has never been observed previously. As the RT lifetime ($\tau_{RT}$) is important to understand the tunnelling behaviour of electrons through the TMO QW, we have computed $\tau_{RT}$ using the energy uncertainty condition at the energy corresponding to the RT, *i.e.* $\tau_{RT} = \hbar/2\Delta E$, where $\hbar = h/2\pi$, $h$ is the Planck's constant, and $\Delta E$ is the half-width-at-half-maximum of the resonant peak[36]. Due to the broad feature of the resonance peak (~1 eV) for our TMO QW, we obtain a rather small $\tau_{RT}$ value (~0.7 fs), which is orders of magnitude smaller than that obtained in semiconductor QWs[36]. This value is also rather small compared to the traverse lifetime across the tunnelling barrier, and therefore, strong coherence is not expected in the complex oxide QWs. In fact, the small $\tau_{RT}$ in our QW SL is expected as such a heterostructure has a rather small relaxation time ($\tau_\mu$ = ~42.4 fs)[24] of the charge carriers, which is also orders of magnitude smaller than those obtained from conventional semiconductors, possibly due to strong correlation[37]. In addition, the peak-to-valley current ratio (PVCR) and the peak current density (PCD) of our TMO-based RT diode are about 1.3 and 120 A/cm$^2$, respectively, at the lowest $T$ (4 K). PVCR is rather small compared to that of conventional semiconductor RT diodes where the typical value is larger than 3 at room $T$. On the other hand, PCD of semiconductor RT diodes span seven orders of magnitude from tens of mA/cm$^2$ to hundreds of kA/cm$^2$. Therefore, the PCD value we have measured from our TMO RT device is within the range of the semiconductor-based RT diodes.

The RT behaviour in our oxide QW SL enhances yet another advantageous functionality, which is unique to the TMOs. Indeed, the resistance ratio between the HRS and LRS is observed to be largely enhanced due to the RT feature. Figure 4 shows the junction resistance of the HRS and LRS measured at 0.5 and –0.5 V, respectively, as a function of $T$. The inset shows the resistance plot as a function of applied bias voltage. The difference between the HRS and LRS resistance increased with decreasing $T$. Below 150 K, we observed the ON/OFF ratio to be larger than $10^5$, while it was slightly larger than 10 at room $T$. Typically, the ON/OFF ratio for bipolar RS is about $10^2$, much smaller than that for unipolar RS. The completely different $T$-dependent behaviour of the resistance (HRS shows insulating behaviour while LRS shows metallic *T*-dependent behaviour.) mainly due to the RT feature greatly enhances the ON/OFF ratio. Below 120 K, the compliance current played a role, somewhat decreasing the HRS resistance value (empty symbols in red). Nevertheless, we could still observe a large resistance ratio (~$10^5$) with an increasing trend towards the lowest *T* we employed. Undoubtedly, such a large enhancement in the resistance ratio stems from the RT behaviour of the QW SL. The drastically enhanced tunnelling probability near the resonant bias voltage substantially decreased the resistance of the LRS near $V_R$. The dip feature for the resistance value at ~1 V for the LRS shown in the inset of Fig. 4 clearly signifies the enhanced tunnelling probability.

In summary, we have observed an intriguing NDR feature in a $SrTiO_3/LaTiO_3/SrTiO_3$ QW superlattice at low temperatures. The NDR behaviour is attributed to a resonant tunnelling occurring through the deliberately designed oxide QWs. In particular, because of the existence of the resonant tunnelling, a largely enhanced ON/OFF ratio has been achieved in the bipolar resistance switching, which occurs at the interface between the heterostructure and metallic substrate. Our study also demonstrates the potential of oxide heterostructures for a quantum mechanical behaviour that has been thought to be possible only in conventional semiconductor heterostructures. Thus, we believe that the discovery of resonant tunnelling through oxide-based QWs can lay down a stepping stone to oxide electronics.

## Methods

**Sample fabrication.** A $[(LaTiO_3)_1/(SrTiO_3)_6]_{10}$ SL sample was fabricated at 700 °C in $10^{-5}$ Torr of oxygen partial pressure using pulsed laser epitaxy. A KrF excimer laser ($\lambda = 248$ nm) with a laser fluence of ~1 J/cm$^2$ was used for ablation of a single crystal STO and a sintered $La_2Ti_2O_7$ target. An atomically flat $TiO_2$ layer terminated (001) Nb-doped STO (0.05%wt) single crystal buffered with 10 u.c. of STO was used as a substrate. The 10 u.c. STO buffer layer was used to ensure a good surface and interface quality of the oxide QW. The growth process was monitored using reflection high-energy electron diffraction (RHEED), ensuring the layer-by-layer growth. Supplementary Figure 1a shows an oscillation of the RHEED specular spot intensity as a function of growth time. Six u.c. layers of STO and one u.c. layer of LTO were consecutively deposited. The *in-situ* RHEED pattern along the [100] direction of the substrate before the SL growth (Supplementary Figure 1b) is clearly maintained even after the growth (Supplementary Figure 1c), indicating an optimized 2D layer-by-layer growth. The surface topographic image by atomic force microscopy (The background of Supplementary Figure 1d) also presents a well-conserved ~0.4 nm high step-and-terrace structure of the substrate even after the deposition of the SL, indicating atomically well-defined surface and interface. After the growth, the SL sample was *in situ* post-annealed for ten min at the growth temperature in a relatively high-oxygen pressure ambience($10^{-2}$ Torr) and then cooled in the same pressure to fully compensate potential oxygen vacancies. We note that the low-pressure growth and subsequent post-annealing are necessary steps for the growth of the perovskite $LaTiO_3$, as it has a very narrow growth window for oxygen pressure. The growth at higher oxygen pressure results in $La_2Ti_2O_7$ or impurity phases. X-ray diffraction (XRD) $\theta$-$2\theta$ scan demonstrated that the designed SL structure was realized (Supplementary Figure 1d). Excellent crystallinity was confirmed from the rocking curve scan as shown in Supplementary Figure 1e (e.g. 006 peak of the SL). The full-width-at-half-maximum value was comparable to that of the Nb-doped substrate. Reciprocal space mapping around the 114 Nb:STO substrate reflection confirms the fully-strained SL sample (Supplementary Figure 1f).

**Electrical characterisation.** The top electrode Pt (~300 $\mu$m in diameter) was patterned by RF sputtering on top of the sample surface using a shadow mask. External bias was applied to the Nb:STO substrate, which served as the bottom electrode. The *T*-dependent current-voltage ($I(V)$) curve was measured using a physical property measurement system (Quantum Design Inc.) with a source measure unit (Kethley 236). The current compliance (0.1 A) and the maximum sweep voltage value (-2 ~ 4 V) were set to prevent the device from damaging.

**Theoretical calculation.** The shape of electrostatic potential (Fig. 1b) induced by La $\delta$-doping in STO was obtained by solving the Poisson and the Schrödinger equation self-consistently, without any electric field applied. The self-consistent calculations were performed iteratively using Broyden's second method until the convergence of the electrostatic potential is reached.[38] An external electric field was then applied to obtain envelope wavefunctions and energy levels of electrons (Fig. 1c). We examined $\epsilon_r$ values of STO in the range of 10-10$^3$, (Note, while $\epsilon_r$ of bulk STO at low $T$ is known to be very large, $\epsilon_r$ strongly varies with the electric field, temperature, and sample geometry.) and effective mass of 4.4 (out-of-plane band effective mass of STO) was used for the calculations.[39]

Acknowledgements

We thank In Rok Hwang, Taekjib Choi, Cheol Seong Hwang and Shinbuhm Lee for valuable discussions. This work was supported by the U.S. Department of Energy, Office of Science, Basic Energy Sciences, Materials Sciences and Engineering Division (W.S.C. and H.N.L.). This research was in part supported by Basic Science Research Program through the National Research Foundation of Korea (NRF) funded by the Ministry of Science, ICT and future Planning (NRF-2014R1A2A2A01006478, W.S.C.) and by the Ministry of Education (NRF-2013R1A1A2057523, S.A.L.). S.L. was supported by KIST Institutional Program (Project No. 2E25440).

Figure legends

Figure 1 | Resonant tunnelling from a quantum oxide superlattice. a, Schematic diagram of a $\delta$-doped quantum oxide superlattice. The $\delta$-doped La layer creates a 2D electron gas within the layer forming a QW structure. b, c, Calculated potential wells (thin black lines) and probability functions (thin lines) of finding electrons, *i.e.*, absolute square of wavefunction, at their corresponding energy levels. At $V = 0$ (b), the wavefunctions with the same energy levels are aligned for each QW. When $V > 0$ the position of wavefunctions are shifted and no longer aligned. At $V = V_R$ (c), the ground state wavefunction of a QW becomes aligned with the first excited state wavefunction in the adjacent QW, allowing a large tunnelling current to flow across the superlattice junction. The dotted line indicates the electric field, which corresponds to $V_R = 1.23$ V for the whole superlattice.

Figure 2 | Resonant tunnelling and negative differential resistance. a, $T$-dependent $I(V)$ characteristics of a superlattice revealing a resonant tunnelling feature superimposed with the resistance switching. The arrows indicate the direction of voltage sweep. The inset shows a similar $T$-dependent current as a function of voltage for a bare Nb:SrTiO$_3$ substrate. b, Differential curves ($dI/dV$) for the same data.

Figure 3 | Theoretical calculation of the confined energy levels and $V_R$ in an oxide QW structure. Effect of $\epsilon_r$ on a, energy levels of confined electrons in the QWs without an external electric field and b, $V_R$ in a $[(LaTiO_3)_1/(SrTiO_3)_6]_{10}$ oxide superlattice. When $\epsilon_r = 100$, we obtain a single resonant condition at $V_R = \sim 1.23$ V, which is consistent with the experimental result.

Figure 4 | *T*-dependent resistance switching. The resistance values for high and low resistance states (i.e., ON and OFF) at -0.5 and 0.5 V are summarized as red and blue filled symbols, respectively. The empty symbols denote the resistance values for high resistance states at -0.5 V with partial switching due to the compliance current limit. The inset shows the $T$-dependent resistance as a function of applied voltage. See the legend in Fig. 2 for temperature indexing.

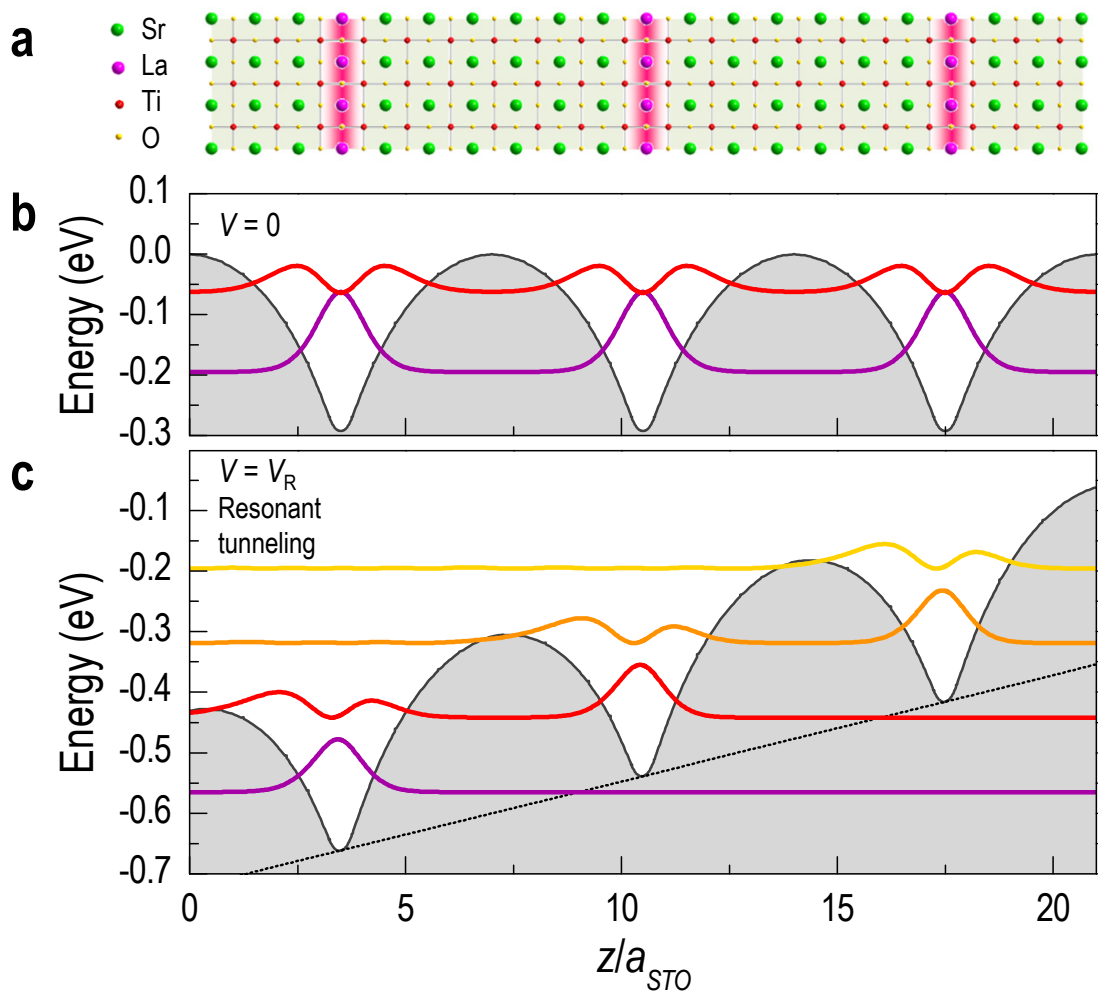

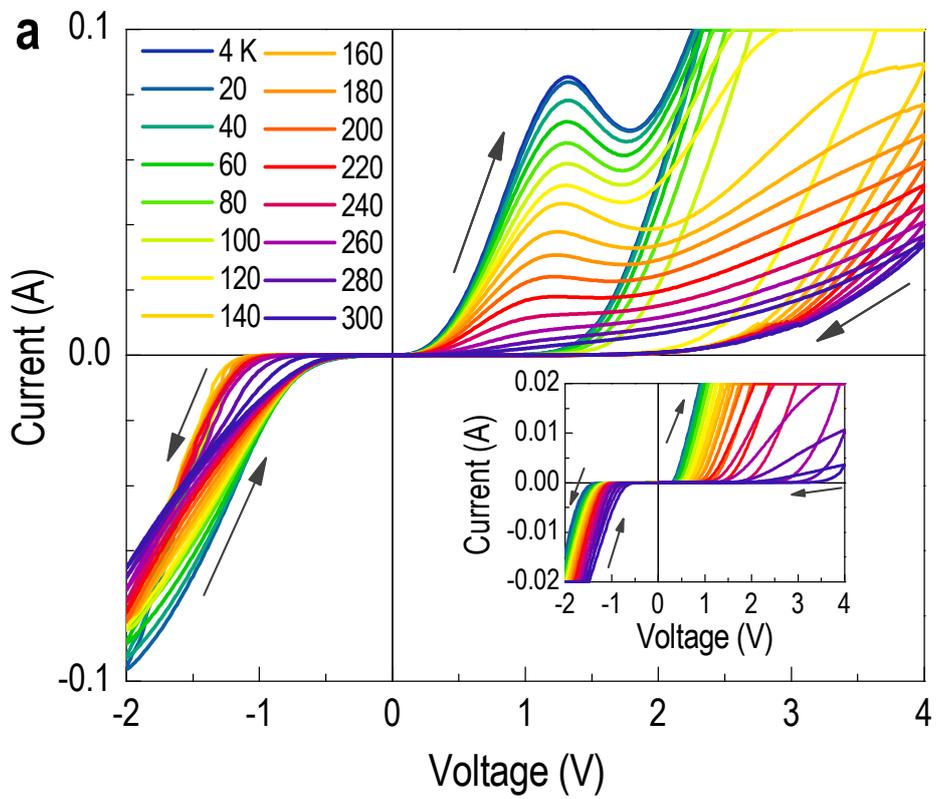

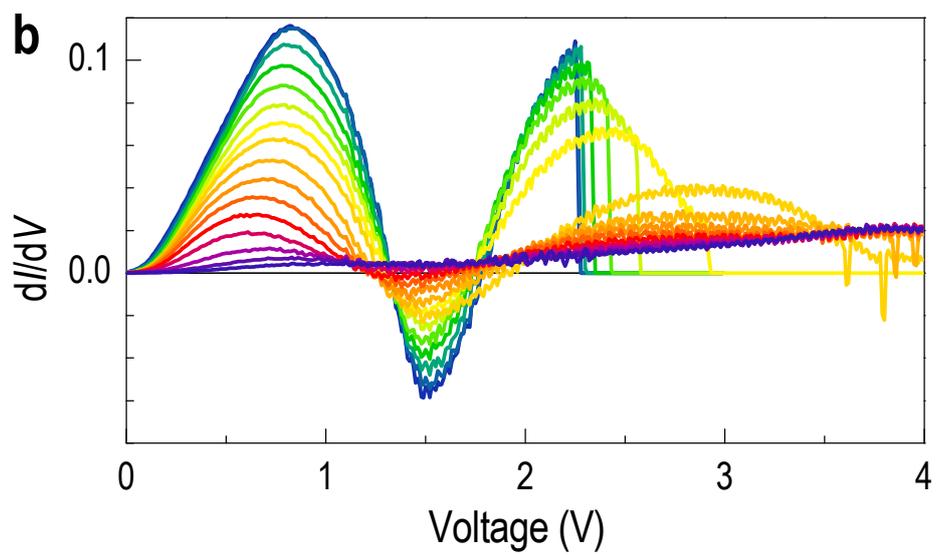

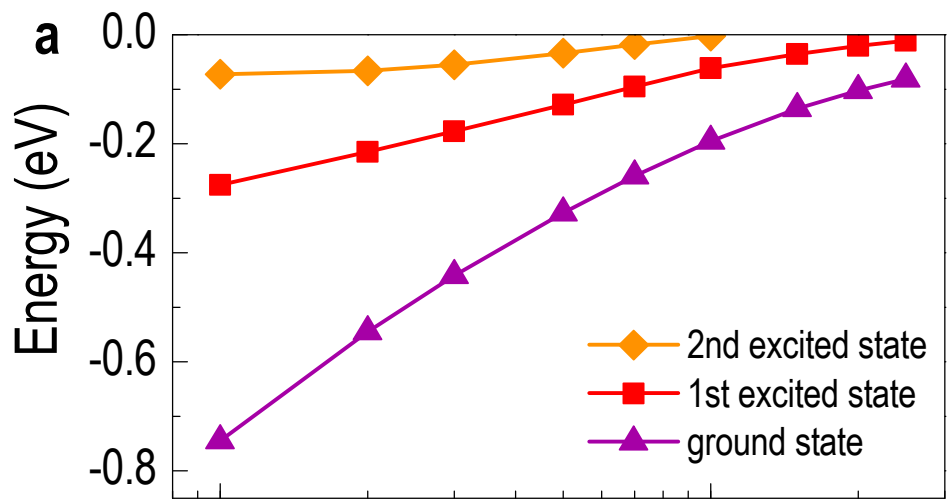
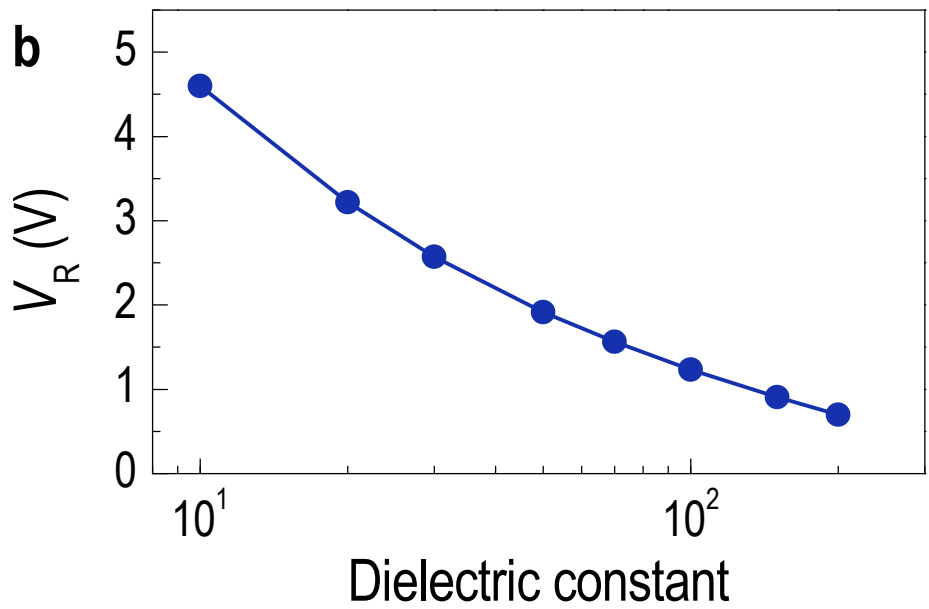

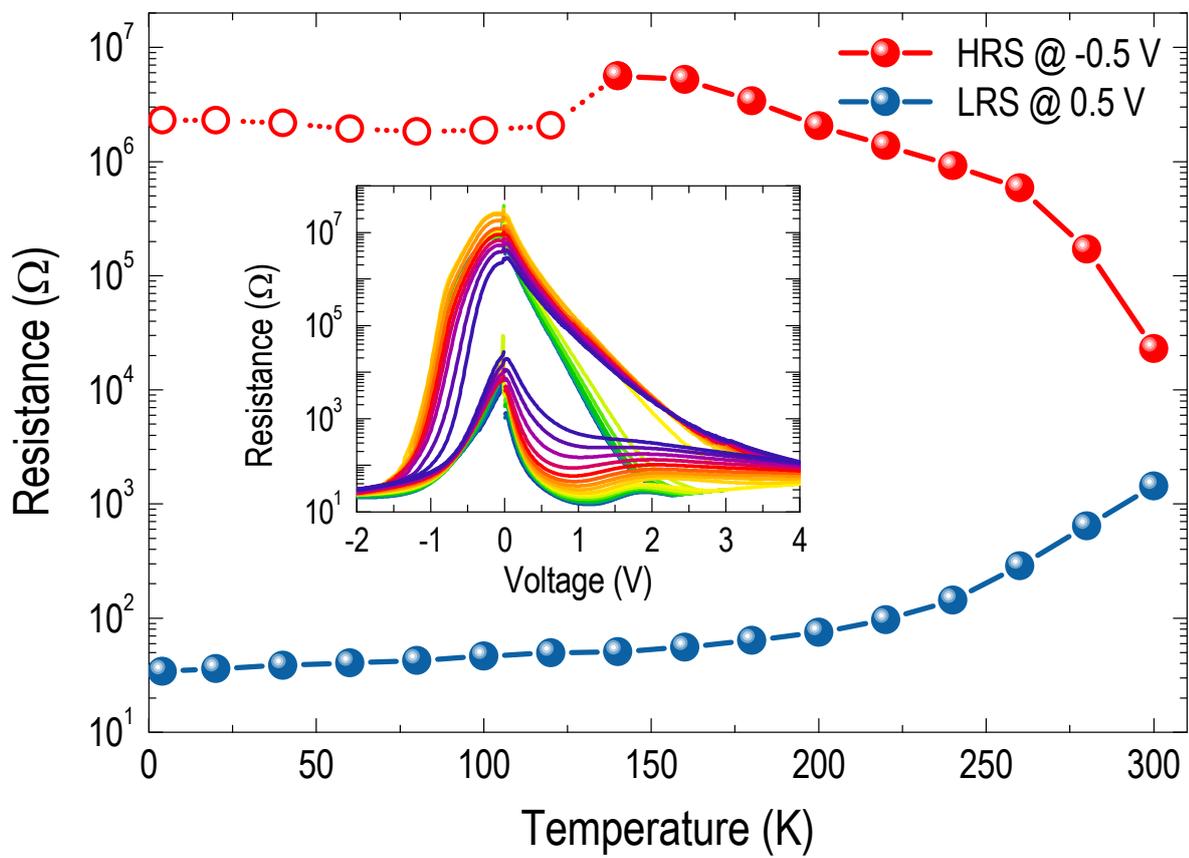

**Supplementary Figures**

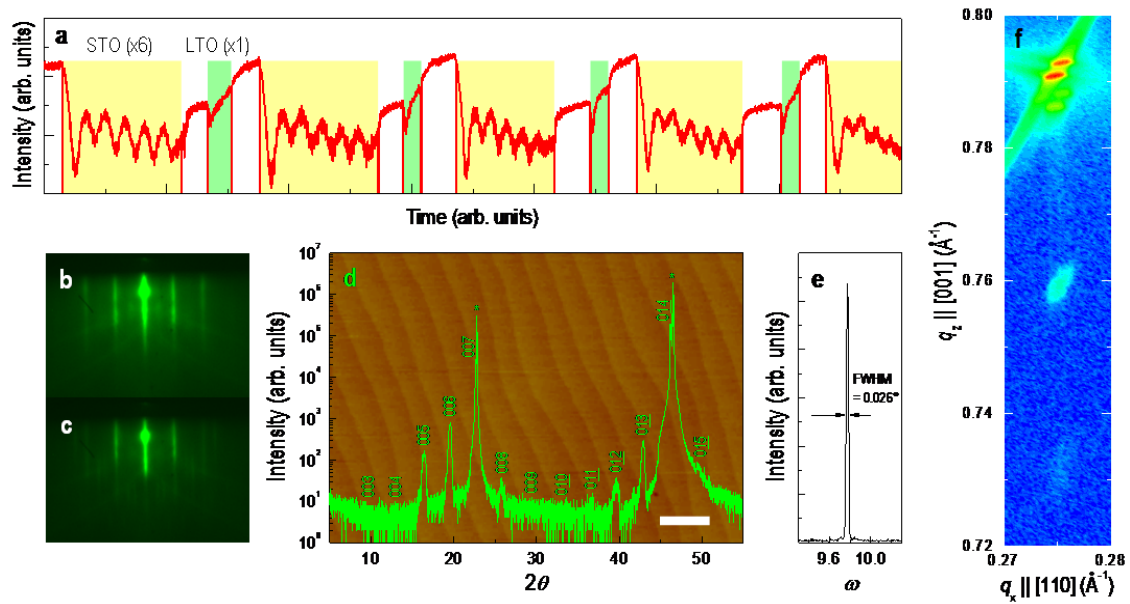

**Supplementary Figure 1 | Fabrication and structural property of a LaTiO$_3$/SrTiO$_3$ superlattice with a well-defined quantum-well structure. a**, RHEED Intensity oscillations during the growth. Bright yellow- and green-shaded regions represent oscillations recorded while growing 6 u.c. of SrTiO$_3$ and 1 u.c. of LaTiO$_3$, respectively. RHEED pattern for **b**, a bare Nb:SrTiO$_3$ substrate before the growth and **c**, after the superlattice growth. 2D layer-by-layer growth is well-maintained throughout the growth. **d**, XRD $\theta$-$2\theta$ scan of the superlattice, clearly showing superlattice satellite peaks for intended structure. The background shows a corresponding topographic image by atomic force microscopy with atomically well-defined single u.c. step-terrace structure conserved even after the growth. The scale bar corresponds to 500 nm. **e**, Rocking curve scan for the 006 peak showing excellent crystallinity. The full-width-at-half-maximum (FWHM) value of 0.026° is comparable to that from a Nb:SrTiO$_3$ substrate. **f**, X-ray reciprocal space map of the superlattice around the substrate SrTiO$_3$ 114 Bragg reflection, demonstrating both a high quality structure and a fully strained state.

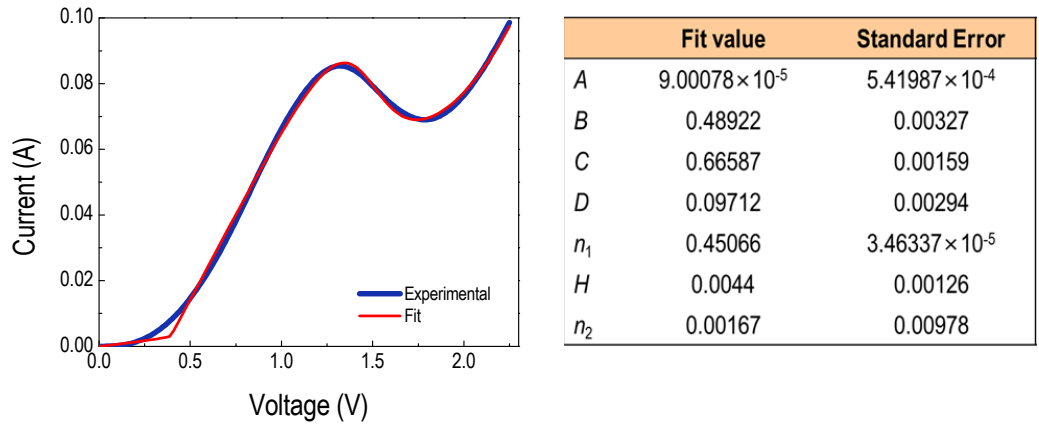

**Supplementary Figure 2 | *I*(*V*) curve simulation.** An excellent fitting result is obtained using analytical model to simulate the I(V) characteristics of the resonant tunneling behaviour of our oxide superlattice.

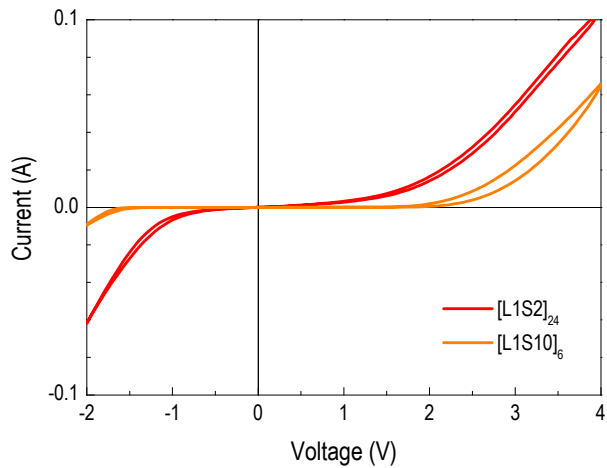

**Supplementary Figure 3 | *I*(*V*) curves for [(LaTiO$_3$)$_1$(SrTiO$_3$)$_2$]$_{24}$ ([L1S2]$_{24}$) and [(LaTiO$_3$)$_1$(SrTiO$_3$)$_{10}$]$_6$ ([L1S10]$_6$) superlattices.** Both samples do not show any NDR behaviour at 20 K, indicating absence of the resonant states. [L1S2]$_{24}$ can be considered as 3D bulk metal, while the [L1S10]$_6$ has too wide barrier for tunneling.

**Supplementary Notes**

**Supplementary Note 1: Simulation of the *I*(*V*) curve using analytic expression based on resonant tunneling.**

An analytic model based on the transmission coefficient in the resonant tunneling devices is used to simulate the *I*(*V*) characteristic of our transition metal oxide superlattices. In particular,

$$J(V) = \frac{em^*kT\Gamma}{4\pi^2\hbar^3} \ln\left[\frac{1+e^{(E_F-E_r+\frac{eV}{2})/kT}}{1+e^{(E_F-E_r-\frac{eV}{2})/kT}}\right]\left[\frac{\pi}{2}+\tan^{-1}\left(\frac{E_r-\frac{eV}{2}}{\frac{\Gamma}{2}}\right)\right], \quad (1)$$

can be used, where $E_F$, $E_r$, and $\Gamma$, is the Fermi energy, the energy of the resonant level, and the resonant width (See the following references for more detail)[1,2]. With the inclusion of the valley current, the above equation can be approximated as[2],

$$J(V) = A \ln\left[\frac{1+e^{(B-C+n_1V)e/kT}}{1+e^{(B-C-n_1V)e/kT}}\right]\left[\frac{\pi}{2}+\tan^{-1}\left(\frac{C-n_1V}{D}\right)\right] + H\left(e^{\frac{n_2eV}{kT}}-1\right). \quad (2)$$

The result of curve fitting is shown in Supplementary Figure 2. An excellent match between the experiment and simulation is shown with the parameters summarized in the table next to the Figure. Particularly, the level splitting between the resonant states (*C* − *B*) could be approximated as 0.1767 eV, which is close to that obtained using the theoretical calculation (0.1334 eV) shown in Fig 3, considering the simplicity of the analytical model.

**Supplementary Note 2: Modification of superlattice period.**

In order to analyse the resonant tunnelling behaviour in further detail, we have modified the period of the superlattices. We have fabricated [(LaTiO$_3$)$_1$(SrTiO$_3$)$_2$]$_{24}$ ([L1S2]$_{24}$) and [(LaTiO$_3$)$_1$(SrTiO$_3$)$_{10}$]$_6$ ([L1S10]$_6$) in addition to the [L1S6]$_{10}$ superlattice which shows a clear NDR feature as shown in the main text. The repetition numbers are selected to roughly conserve the total superlattice thickness. Supplementary Figure 3 shows the *I*(*V*) characteristics of the superlattices measured at 20 K. Both superlattices do not show any NDR behaviour and only a weak hysteresis is present. The reasons for the absence of resonant states in the two superlattices are the following. [L1S2]$_{24}$ superlattice can be considered a 3D metal as the charges from the two adjacent LaTiO$_3$ layers overlap in the SrTiO$_3$ layer. Recent theoretical calculation[5] and also the scanning transmission electron microscopy – electron energy loss spectroscopy[6] suggest the extent of the charges into the SrTiO$_3$ layer is about 3 unit cells, which is larger than the spacing designed here. Therefore, the quantum well structure cannot be defined in [L1S2]$_{24}$ superlattice. On the other hand, [L1S10]$_6$ superlattice has much thicker barrier width compared to [L1S6]$_{10}$ superlattice. Therefore, the overlapping of the wavefunction and hence the tunnelling probability drastically decreases. Indeed the sample becomes more insulating.

**Supplementary Note 3: Discussion on the phase transitions of the constituent layers**

Perovskite SrTiO$_3$ and LaTiO$_3$ undergo important phase transitions with lowering the temperature. SrTiO$_3$ undergoes a structural transition from a cubic to a tetragonal phase at around 105 K and LaTiO$_3$ undergoes a phase transition from a nonmagnetic insulator to an antiferromagnetic insulator near 146 K.[3, 4] While these phase transitions might influence the RT behaviour of the TMO superlattices, we did not find any anomaly across the temperatures within the error bar of our experiment. It should be noted that more detailed structural and magnetic studies across the temperatures might elucidate the subtle correlation between the phase transition in oxide perovskites and quantum tunneling behaviour.